\documentclass[aps,prc,twocolumn,floatfix,nofootinbib,showpacs,superscriptaddress]{revtex4-1}

\usepackage{amsmath,amsfonts,amssymb,bm}
\usepackage{graphicx}
\usepackage{color}
\usepackage{pzccal}
\usepackage{soul}
\definecolor{purple}{rgb}{0.5,0,0.5}
\definecolor{blue}{rgb}{0.0,0,0.9}

\begin{document}

\title{Insights into the $\gamma^\ast N \to \Delta$ transition}

\author{Jorge Segovia}
\affiliation{Physics Division, Argonne National Laboratory, Argonne, Illinois 60439, USA}

\author{Chen Chen}
\affiliation{Institute for Theoretical Physics and Department of Modern Physics,
University of Science and Technology of China, Hefei 230026, P.\ R.\ China}

\author{Craig D.~Roberts}
\affiliation{Physics Division, Argonne National Laboratory, Argonne, Illinois 60439, USA}

\author{Shaolong Wan}
\affiliation{Institute for Theoretical Physics and Department of Modern Physics,
University of Science and Technology of China, Hefei 230026, P.\ R.\ China}

\date{28 August 2013}

\begin{abstract}
The $\gamma^\ast N \to \Delta(1232)$ transition is a window on hadron shape deformation, the applicability of perturbative QCD at moderate momentum transfers, and the influence of nonperturbative phenomena on hadronic observables.  We explain that the Ash-convention magnetic transition form factor must fall faster than the neutron's magnetic form factor and nonzero values for the associated quadrupole ratios reveal the impact of quark orbital angular momentum within the nucleon and $\Delta(1232)$; and show that these quadrupole ratios do approach their predicted asymptotic limits, albeit slowly.
\end{abstract}

\pacs{
13.40.Gp, 	
14.20.Dh,	
14.20.Gk,   
12.38.Aw    
}

\maketitle

\hspace*{-\parindent}\textbf{I.~Introduction}.
%
%
The $\Delta(1232)$ family were the first resonances discovered in $\pi N$ reactions \cite{Fermi:1952zz,Anderson:1952nw,Nagle:1984sg}.  Through extensive study we have come to know that $\Delta(1232)$-baryons are positive parity, isospin $I=\frac{3}{2}$, total-spin $J=\frac{3}{2}$ bound-states with no net strangeness \cite{Beringer:1900zz}.  As such, the $\Delta^+$ and $\Delta^0$ can be viewed, respectively, as isospin- and spin-flip excitations of the proton and neutron.
Since pions are a complex probe, it is sensible to exploit the relative simplicity of virtual photons in order study the $\Delta$-resonance's structure; viz., through the transitions $\gamma^\ast N \to \Delta$.  This is possible at
intense, energetic electron-beam facilities; and data on the $\gamma^\ast p \to \Delta^+$ transition are now available for $0 \leq Q^2 \lesssim 8\,$GeV$^2$ \cite{Aznauryan:2011ub,Aznauryan:2011qj}.

The $\gamma^\ast p \to \Delta^+$ data has stimulated much theoretical analysis, and speculation about, \emph{inter alia}:
the relevance of perturbative QCD (pQCD) to processes involving moderate momentum transfers \cite{Carlson:1985mm,Pascalutsa:2006up,Aznauryan:2011qj};
shape deformation of hadrons \cite{Alexandrou:2012da};
and the role that resonance electroproduction experiments can play in exposing nonperturbative features of QCD, such as the nature of confinement and dynamical chiral symmetry breaking (DCSB) \cite{Aznauryan:2012ba}.

The \mbox{$N\to\Delta$} transition is described by three form factors \cite{Jones:1972ky}: magnetic-dipole, $G_M^\ast$; electric quadrupole, $G_E^\ast$; and Coulomb (longitudinal) quadrupole, $G_C^\ast$.  They arise through consideration of the $N\to \Delta$ transition current:
\begin{equation}
J_{\mu\lambda}(K,Q) =
\Lambda_{+}(P_{f})R_{\lambda\alpha}(P_{f})i\gamma_{5}\Gamma_{\alpha\mu}(K,Q)\Lambda_
{+}(P_{i}),
\label{eq:JTransition}
\end{equation}
where: $P_{i}$, $P_{f}$ are, respectively, the incoming nucleon and outgoing $\Delta$ momenta, with $P_{i}^{2}=-m_{N}^{2}$, $P_{f}^{2}=-m_{\Delta}^{2}$; the incoming photon momentum is $Q_\mu=(P_{f}-P_{i})_\mu$ and $K=(P_{i}+P_{f})/2$; and $\Lambda_{+}(P_{i})$, $\Lambda_{+}(P_{f})$ are, respectively, positive-energy projection operators for the nucleon and $\Delta$, with the Rarita-Schwinger tensor projector $R_{\lambda\alpha}(P_f)$ arising in the latter connection.  (Our Euclidean metric conventions are described, e.g., in App.\,A of Ref.\,\protect\cite{Chen:2012qr}.)

In order to succinctly express $\Gamma_{\alpha\mu}(K,Q)$, we define
%
$\hat K_{\mu}^{\perp} = {\cal T}_{\mu\nu}^{Q} \hat K_{\nu}
= (\delta_{\mu\nu} - \hat Q_\mu \hat Q_\nu) \hat K_\nu$,
$\hat K^2 = 1= \hat Q^2$, in which case, with $\mathpzc{k} = \sqrt{(3/2)}(1+m_\Delta/m_N)$,
$\varsigma = Q^{2}/[2\Sigma_{\Delta N}]$,
$\lambda_\pm = \varsigma + t_\pm/[2 \Sigma_{\Delta N}]$
where $t_\pm = (m_\Delta \pm m_N)^2$,
$\lambda_m = \sqrt{\lambda_+ \lambda_-}$,
$\Sigma_{\Delta N} = m_\Delta^2 + m_N^2$, $\Delta_{\Delta N} = m_\Delta^2 - m_N^2$,
\begin{eqnarray}
\nonumber \Gamma_{\alpha\mu}(K,Q) & =&
\mathpzc{k} \left[\frac{\lambda_m}{2\lambda_{+}}(G_{M}^{\ast}-G_{E}^{\ast})\gamma_{5}
\varepsilon_{\alpha\mu\gamma\delta} \hat K_{\gamma}\hat{Q}_{\delta}  \right. \\
&&
\left.  \hspace{0.25cm}
- G_{E}^{\ast}
{\cal T}_{\alpha\gamma}^{Q}
{\cal T}_{\gamma\mu}^{K}
- \frac{i\varsigma}{\lambda_m}G_{C}^{\ast}\hat{Q}_{\alpha} \hat K^\perp_{\mu}\right].
\label{eq:Gamma2Transition}
\end{eqnarray}

Given the current, one may obtain the form factors using any three sensible projection operations; e.g., with $\mathpzc{d}=\Delta_{\Delta N}/[2 \Sigma_{\Delta N}]$,
$\mathpzc{n}= \sqrt{1-4\mathpzc{d}^{2}}/[4i\mathpzc{k}\lambda_m]$), and
\begin{subequations}
\label{contractions}
\begin{eqnarray}
s_{1} (\mathpzc{d}-\varsigma )&=& \mathpzc{n}  \sqrt{\varsigma(1+2\mathpzc{d})} {\cal T}^{K}_{\mu\nu}\hat K^\perp_{\lambda} {\rm tr} \gamma_{5}J_{\mu\lambda}\gamma_{\nu}\,, \\
s_{2} \lambda_m &=& \mathpzc{n} \lambda_{+} {\cal T}^{K}_{\mu\lambda} {\rm tr} \gamma_{5} J_{\mu \lambda}\,, \\
s_{3} (\mathpzc{d}-\varsigma) \lambda_m &=&  3 \mathpzc{n}  \lambda_+ (1+2\mathpzc{d}) \hat K^\perp_{\mu}\hat K^\perp_{\lambda} {\rm tr}\gamma_{5}J_{\mu\lambda} \,,
\end{eqnarray}
\end{subequations}
then
$G_{M}^{\ast} = 3 [ s_{2}+s_{1}]$, $G_{E}^{\ast} = s_{2}-s_{1}$, $G_{C}^{\ast} = s_{3}$.
%
%


In analyses of baryon electromagnetic properties, using a quark model framework which implements a current that transforms according to the adjoint representation of spin-flavour $SU(6)$, one finds simple relations between magnetic-transition matrix elements \cite{Beg:1964nm,Buchmann:2004ia}:
\begin{equation}\label{eqBeg}
\langle p | \mu | \Delta^+\rangle = -\langle n | \mu | \Delta^0\rangle\,,\;
\langle p | \mu | \Delta^+\rangle = - \surd 2 \langle n | \mu | n \rangle
\end{equation}
i.e., the magnetic components of the $\gamma^\ast p \to \Delta^+$ and $\gamma^\ast n \to \Delta^0$ are equal in magnitude and, moreover, simply proportional to the neutron's magnetic form factor.  Furthermore, both the nucleon and $\Delta$ are $S$-wave states (neither is deformed) and hence $G_{E}^{\ast} \equiv 0 \equiv G_{C}^{\ast}$ \cite{Alexandrou:2012da}.

Equation~\eqref{eqBeg}-left is consistent with pQCD \cite{Carlson:1985mm} in the following sense: both suggest that $G_{M}^{\ast p}(Q^2)$ should decay with $Q^2$ at the same rate as the neutron's magnetic form factor, which is dipole-like in QCD.  It is usually argued that this is not the case empirically \cite{Aznauryan:2011ub,Aznauryan:2011qj}, a claim that stimulated our interest in the $\gamma^\ast N \to \Delta$ transition.

\smallskip

\hspace*{-\parindent}\textbf{II.~General Observations}.
%
Baryon bound-states in quantum field theory may be described by a Faddeev amplitude, $\Psi$, obtained from a Poincar\'e-covariant Faddeev equation \cite{Cahill:1988dx}, which sums all interactions that can take place between the three quarks that define its valence-quark content.  The appearance of nonpointlike colour-antitriplet diquark correlations \cite{Cahill:1987qr,Maris:2002yu} within the proton is a dynamical prediction of Faddeev equation studies; and empirical evidence in support of the presence of diquarks in the proton is accumulating
\cite{Close:1988br,Cloet:2005pp,Wilson:2011aa,Cates:2011pz,Cloet:2012cy}.

As the nucleon and $\Delta$ have positive parity, $J^P=0^+$ (scalar) and $J^P=1^+$ (axial-vector) diquarks are the dominant correlations within them \cite{Roberts:2011cf,Chen:2012qr}.  The presence of pseudoscalar and vector diquarks can be ignored because such correlations are characterised by much larger mass-scales and they have negative parity \cite{Roberts:2011cf,Chen:2012qr}.
Owing to Fermi-Dirac statistics, scalar diquarks are necessarily $I=0$ states, whilst axial-vector diquarks are $I=1$ \cite{Cahill:1987qr}.  The nucleon ground-state contains both $0^+$ and $1^+$ diquarks, whereas the $\Delta(1232)$-baryon contains only axial-vector diquarks because it is impossible to combine an $I=0$ diquark with an $I=1/2$ quark to obtain $I=3/2$.

\begin{figure}[t]
\begin{centering}
\includegraphics[clip,width=0.5\linewidth]{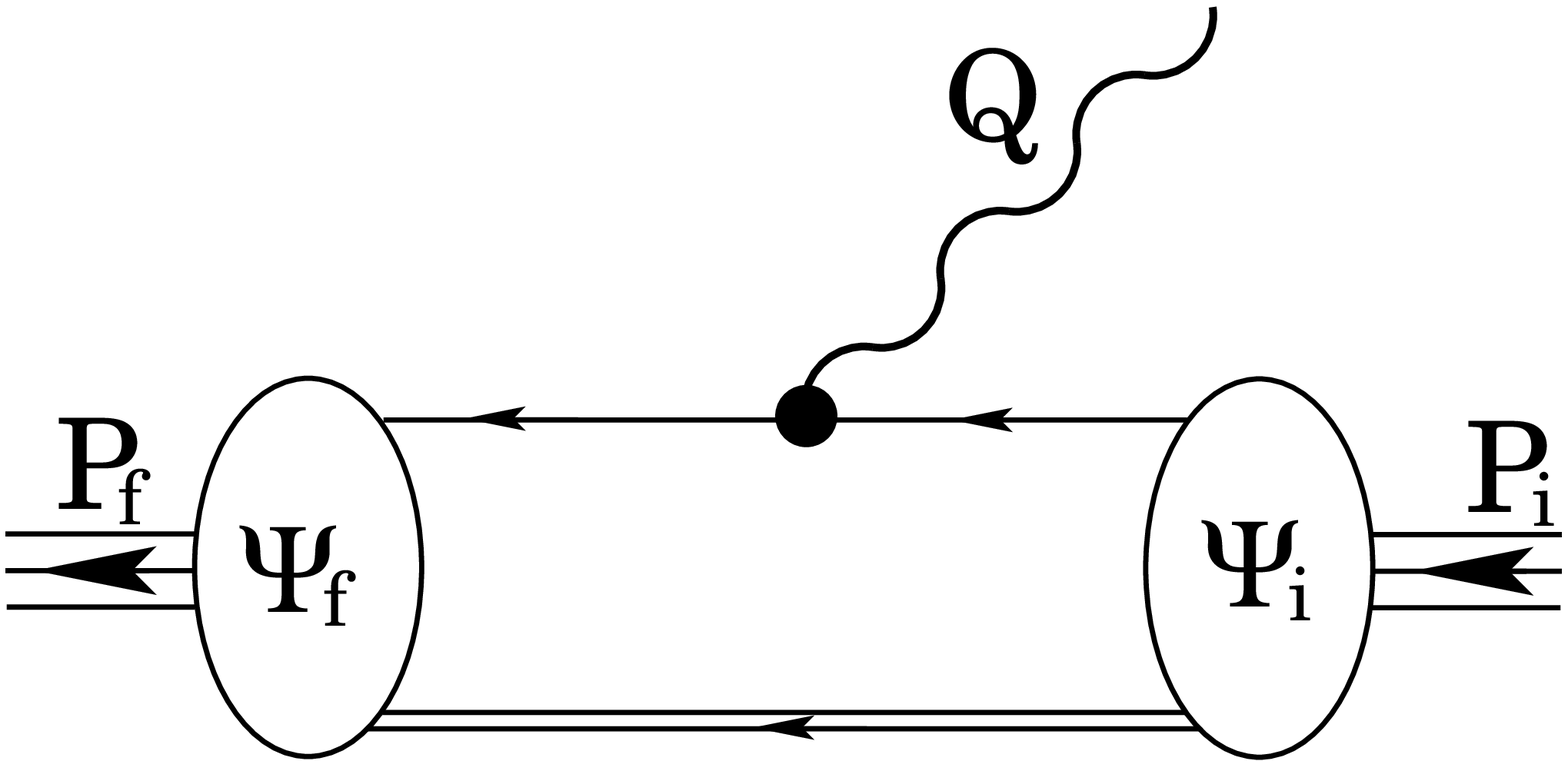}
\vspace*{0.3em}

\includegraphics[clip,width=0.5\linewidth]{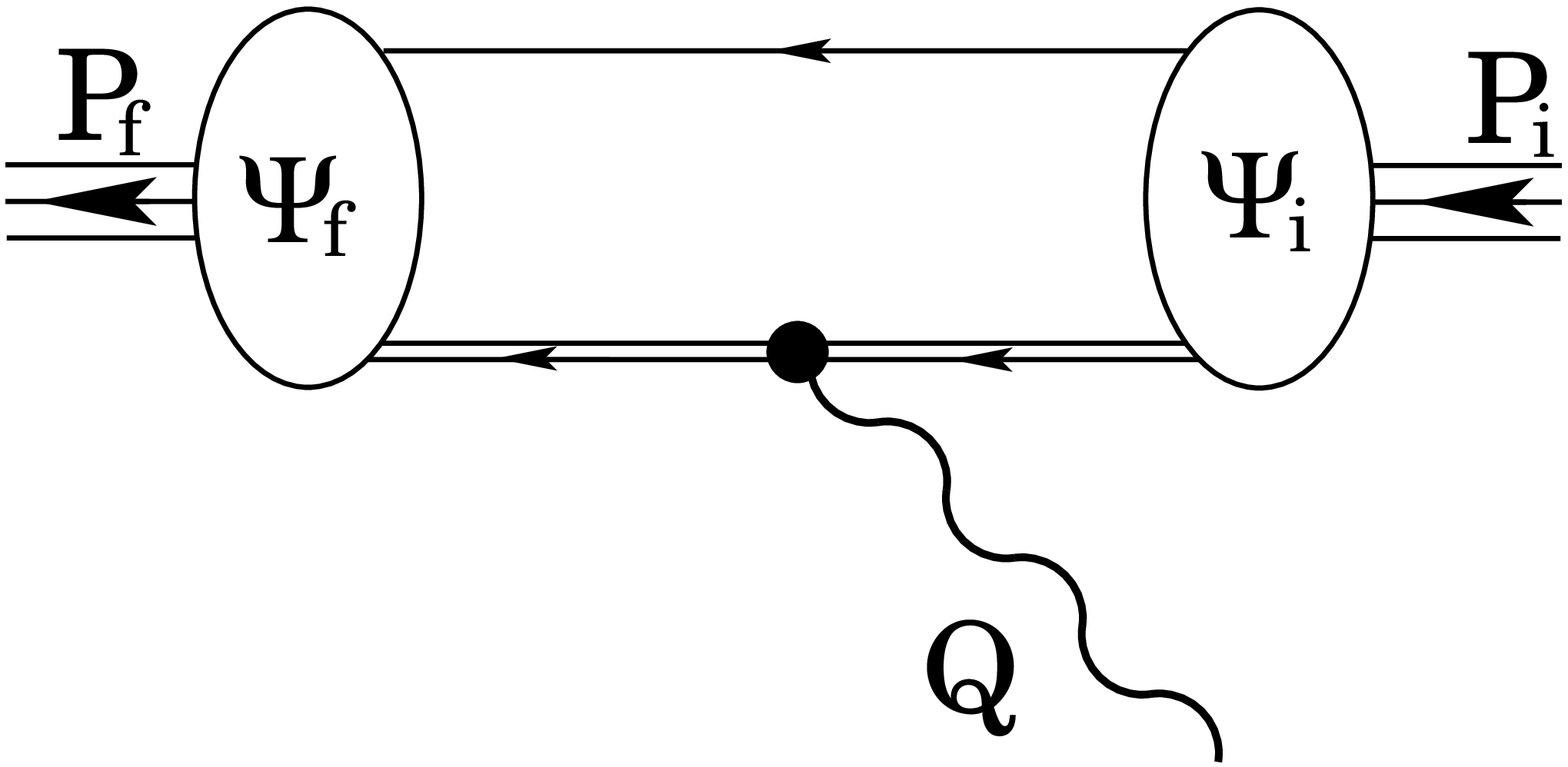}
\vspace*{0.3em}

\includegraphics[clip,width=0.5\linewidth]{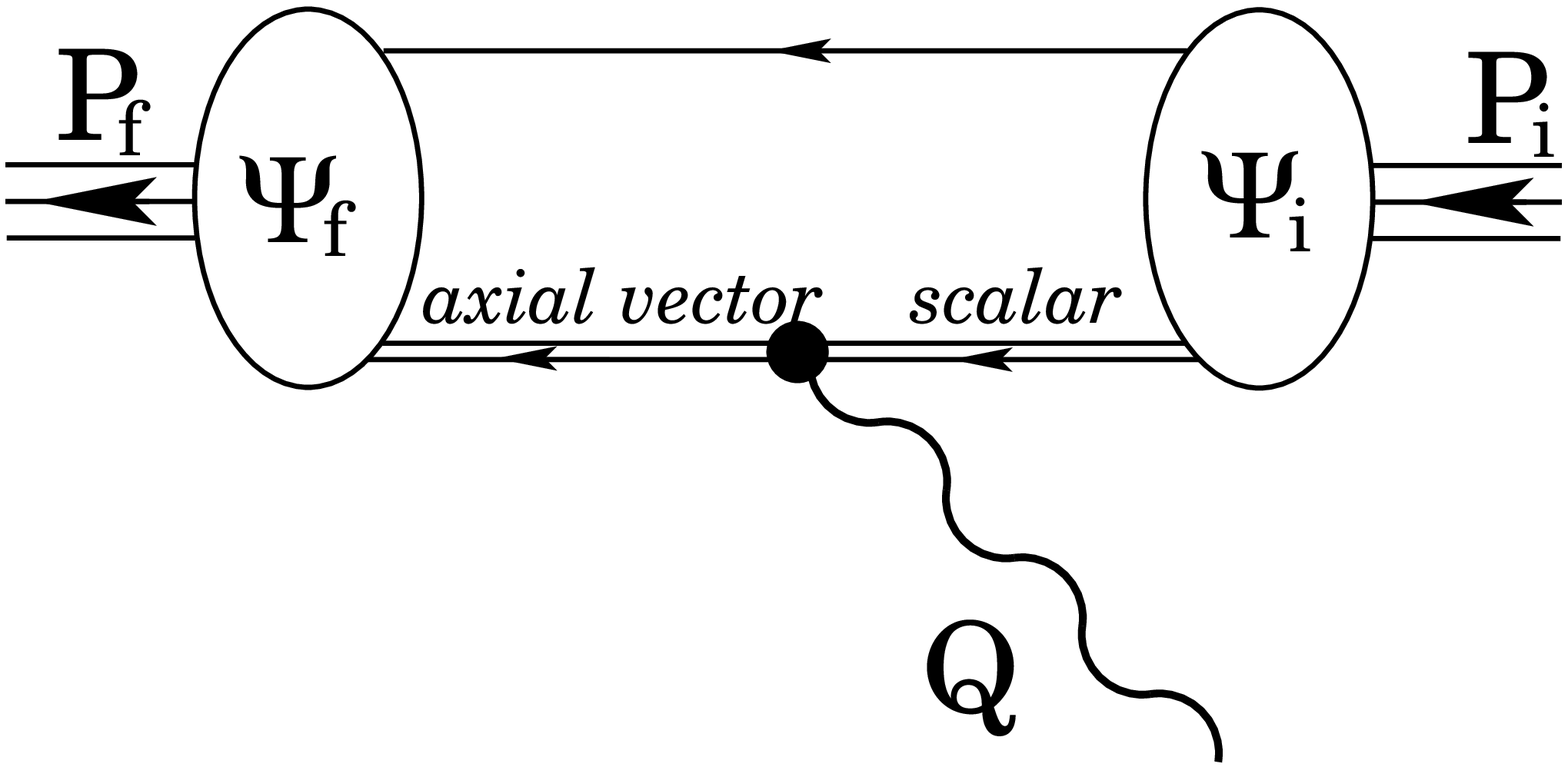}
\end{centering}
\caption{\label{fig:Transitioncurrent} One-loop diagrams in the $N\to \Delta$ vertex:
single line, dressed-quark propagator, $S(p)$; double line, diquark propagator; and vertices, respectively, incoming nucleon, $\Psi_i$, and outgoing $\Delta$, $\Psi_f$.  From top to bottom, the diagrams describe the photon coupling: directly to a dressed-quark; to a diquark, in an elastic scattering event; or inducing a transition between scalar and axial-vector diquarks.
The $\Delta$ resonance contains only axial-vector diquark correlations so here only such diquarks appear in the top and middle diagrams.
In the general case, there are three more diagrams, described in detail elsewhere \protect\cite{Cloet:2008re}.  They represent two-loop integrals.
}
\end{figure}


For baryons constituted as described above, the elastic and transition currents are represented by the diagrams described in association with Fig.\ref{fig:Transitioncurrent}.  Plainly, with the presence of strong diquark correlations, the assumption of $SU(6)$ symmetry for the associated state-vectors and current is invalid.  Notably, too, since scalar diquarks are absent from the $\Delta$, only axial-vector diquark correlations contribute in the top and middle diagrams of Fig.\ref{fig:Transitioncurrent} when one or both of the vertices involves a $\Delta(1232)$-baryon.

Each of the diagrams in Fig.\ref{fig:Transitioncurrent} can be expressed like Eq.\,(\ref{eq:JTransition}), so that we may represent them as
\begin{equation}
\label{GammaTransition}
\Gamma_{\mu\lambda}^{m}(K,Q) = \Lambda_{+}(P_{f})R_{\lambda\alpha}(P_{f}) {\cal
J}_{\mu\alpha}^{n}(K,Q) \Lambda_{+}(P_{i})\,,
\end{equation}
where $m=1,2,\,$\ldots enumerates the diagrams, from top to bottom.  
The top diagram describes a photon coupling directly to a dressed-quark with the axial-vector diquark acting as a bystander.  If the initial-state is a proton, then it contains two axial-vector diquark isospin states $(I,I_z) = (1,1)$, $(1,0)$, with flavour content $\{uu\}$ and $\{ud\}$, respectively: in the isospin-symmetry limit, they appear with relative weighting $(\sqrt{2/3})$:$(-\sqrt{1/3})$, which are just the appropriate isospin-coupling Clebsch-Gordon coefficients.  These axial-vector diquarks also appear in the final-state $\Delta^+$ but with the orthogonal weighting; i.e., $(\sqrt{1/3})$:$(\sqrt{2/3})$. For the process $\gamma^\ast p \to \Delta^+$, Diagram~1 therefore represents a sum, which may be written
\begin{equation}
{\cal J}_{\mu\alpha}^{1 p} =
(\sqrt{2}/3) e_d {\mathcal I}_{\mu \alpha}^{1 \{uu\}}
-(\sqrt{2}/3) e_u {\mathcal I}_{\mu\alpha}^{1\{ud\}},
\label{eqJ1p}
\end{equation}
where we have extracted the isospin and charge factors associated with each scattering.  Plainly, if the $\{uu\}$ diquark is a bystander, then the $d$-quark is the active scatterer, and hence appears the factor $e_d=(-1/3)$.  Similarly, $e_u=2/3$ appears with the $\{ud\}$ diquark bystander.

Now, having extracted the isospin and electric-charge factors, nothing remains to distinguish between the $u$- and $d$-quarks in the isospin-symmetry limit.  Hence,
\begin{eqnarray}
\label{calIisospin}
&& {\mathcal I}_{\mu\alpha}^{1 \{uu\}}(K,Q) \equiv {\mathcal I}_{\mu\alpha}^{1\{ud\}}(K,Q)
=: {\mathcal I}_{\mu\alpha}^{1 \{qq\}}(K,Q) \,,\\
\label{J1pzero}
&\Rightarrow & {\cal J}_{\mu\alpha}^{1 p}(K,Q) = (-\sqrt{2}/3) {\mathcal I}_{\mu\alpha}^{1 \{qq\}}(K,Q)\,.
\end{eqnarray}
It is known that diagrams with axial-vector diquark spectators do not contribute to proton elastic form factors (Eq.\,(C5) in Ref.\,\cite{Wilson:2011aa}), so the analogous contribution is absent from the proton's elastic form factors.  However, this hard contribution is present in neutron elastic form factors.  In general, form factors also receive a hard contribution from the two-loop diagrams omitted in Fig.\ref{fig:Transitioncurrent}.  In proton and neutron elastic magnetic form factors, respectively, the large-$Q^2$ behaviour of this contribution matches that produced by Diagram~1 \cite{Cloet:2008re}.

The remaining two diagrams in Fig.\ref{fig:Transitioncurrent}; i.e., the middle and bottom images, describe a photon interacting with a composite object whose electromagnetic radius is nonzero.  (Indeed \cite{Roberts:2011wy}: $r_{1^+} \gtrsim r_\pi$.)  They must therefore produce a softer contribution to the transition form factors than anything obtained from the top diagram.

It follows from this discussion that the fall-off rate of $G_M^\ast(Q^2)$ in the $\gamma^\ast p \to \Delta^+$ transition must match that of $G_M^n(Q^2)$.
%
With isospin symmetry, Eq.\,\eqref{eqBeg}-left is valid, so the same is true of the $\gamma^\ast n \to \Delta^0$ magnetic form factor.  Note that these are statements about the dressed-quark-core contributions to the transitions.  They will be valid empirically outside that domain upon which meson-cloud effects are important; i.e., for $Q^2\gtrsim 2\,$GeV$^2$ \cite{Sato:2000jf,JuliaDiaz:2006xt}.

\smallskip

\hspace*{-\parindent}\textbf{III.~Quantitative Illustration}.
Since these observations are straightforward, we choose to illuminate them within a simple framework, using a symmetry-preserving Dyson-Schwinger equation (DSE) treatment of a vector$\,\times\,$vector contact-interaction (CI).  A body of recent work \cite{GutierrezGuerrero:2010md,Roberts:2010rn,%
Roberts:2011cf,Roberts:2011wy,Wilson:2011aa,Chen:2012qr,Chen:2012txa} has shown that this framework produces results which, when analysed judiciously, are qualitatively and semi-quantitatively equivalent to those obtained with the most sophisticated interactions thus far employed in the leading-order (rainbow-ladder \cite{Bender:1996bb}) truncation of QCD's DSEs.  Our illustration is therefore representative of that class of studies.

To proceed, we need only adapt the transition form factor formulae in Ref.\,\cite{Wilson:2011aa} to the case of a final-state $\Delta$.  Owing to the interaction's simplicity, there are no two-loop contributions to the form factors, so the diagrams depicted in Fig.\,\ref{fig:Transitioncurrent} are all that need be considered.  Their structure and analysis is explained in the Appendix.

Following Ref.\,\cite{Wilson:2011aa}, our CI is specified by: an interaction strength $\alpha_{\rm IR}=0.93 \pi$; a confinement mass-scale $\Lambda_{\rm ir}=0.24\,$GeV and ultraviolet cutoff $\Lambda_{\rm uv} = 0.905\,$GeV; and a $u=d$ current-quark mass $m=7\,$MeV, which yields a dressed-quark mass $M=0.368\,$GeV via the rainbow gap equation.  With these inputs, one proceeds directly to predictions for the form factors.

\begin{figure}[t]
\begin{centering}
\includegraphics[clip,width=0.70\linewidth]{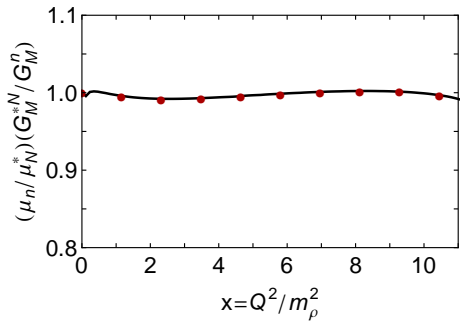}
\end{centering}
\caption{\label{figRatio} \emph{Solid curve} -- $\mu_n G_{M}^{\ast p}/\mu^\ast_p G_{M}^{n}$ as a function of $x=Q^2/m_\rho^2$; and \emph{filled circles} -- $\mu_n G_{M}^{\ast n}/\mu^\ast_n G_{M}^{n}$.
For $N=p,n$, $\mu^\ast_N = G_M^{\ast N}(Q^2=0)$; and $\mu_n = G_M^n(Q^2=0)$.  The elastic form factor results are those presented in Ref.\,\protect\cite{Wilson:2011aa}, so that the comparison is internally consistent.}
\end{figure}

In Fig.\,\ref{figRatio} we compare the momentum-dependence of the magnetic $\gamma^\ast p \to \Delta^+$ and $\gamma^\ast n \to \Delta^0$ form factors with $G_M^n(Q^2)$.  The prediction explained in Sec.\,II is evident in a near identical momentum dependence.

In connection with experiment, our CI treatment of the $N \to \Delta$ transition is quantitatively inadequate for two main reasons.  Namely, a CI which produces Faddeev amplitudes that are independent of relative momentum must underestimate the quark orbital angular momentum content of the bound-state; and the truncation which produces the momentum-independent amplitudes also suppresses the three two-loop diagrams in the current of Fig.\,\ref{fig:Transitioncurrent}.  The detrimental effect can be illustrated via our computed values for the contributions to $G_M^\ast(0)$ that arise from the overlap $1^+$-diquark($\Delta$)$\leftarrow$$1^+$-diquark($N$) cf.\ $1^+$-diquark($\Delta$)$\leftarrow$$0^+$-diquark($N$).  We find $0.85/0.18$, values that may be compared with those in Table~3 of Ref.\,\cite{Eichmann:2011aa}, which uses momentum-dependent DSE kernels: $0.96/1.27$.  One may show algebraically that the omitted two-loop diagrams facilitate a far greater contribution from axial($\Delta$)-scalar($N$) mixing and the presence of additional orbital angular momentum enhances both.

\begin{figure}[t]
\begin{centering}
\includegraphics[clip,width=0.7\linewidth]{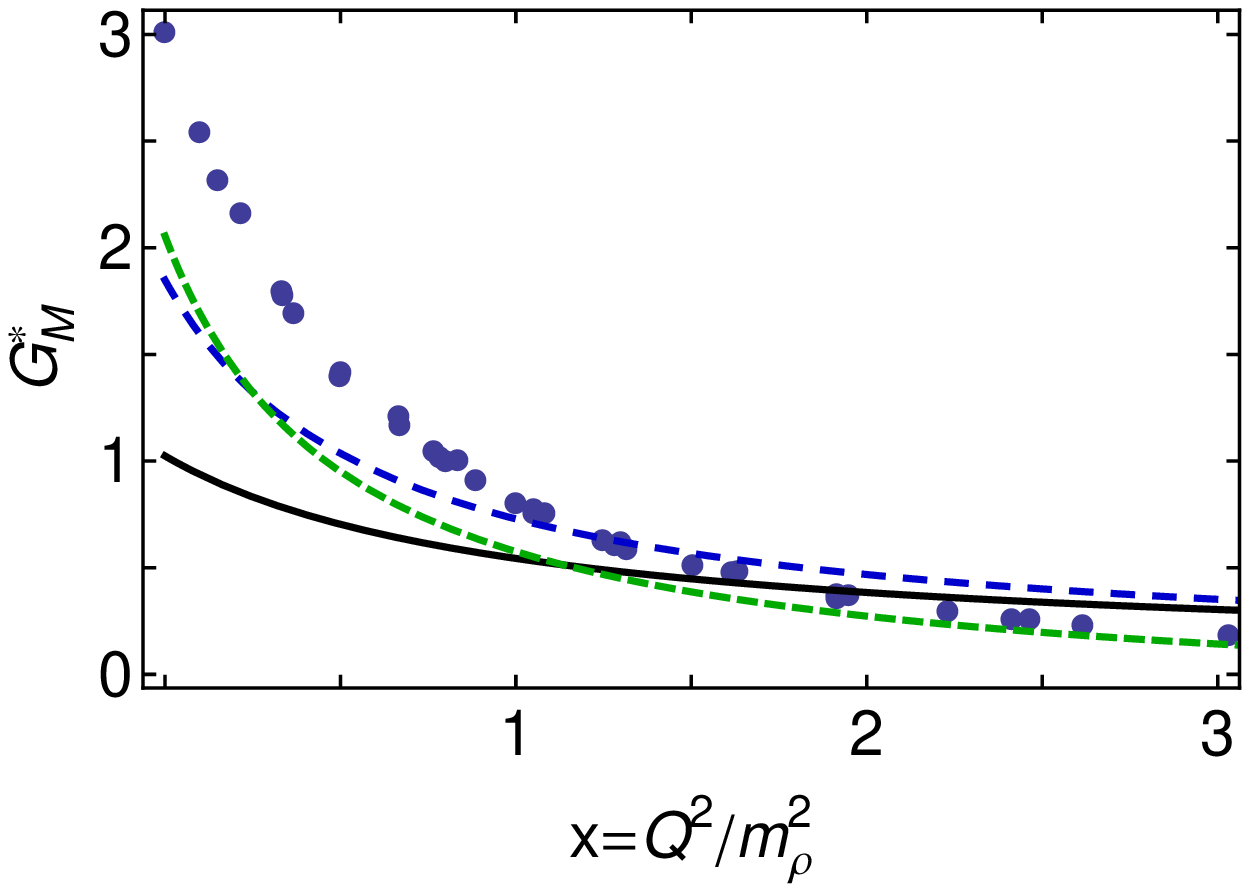}
\includegraphics[clip,width=0.7\linewidth]{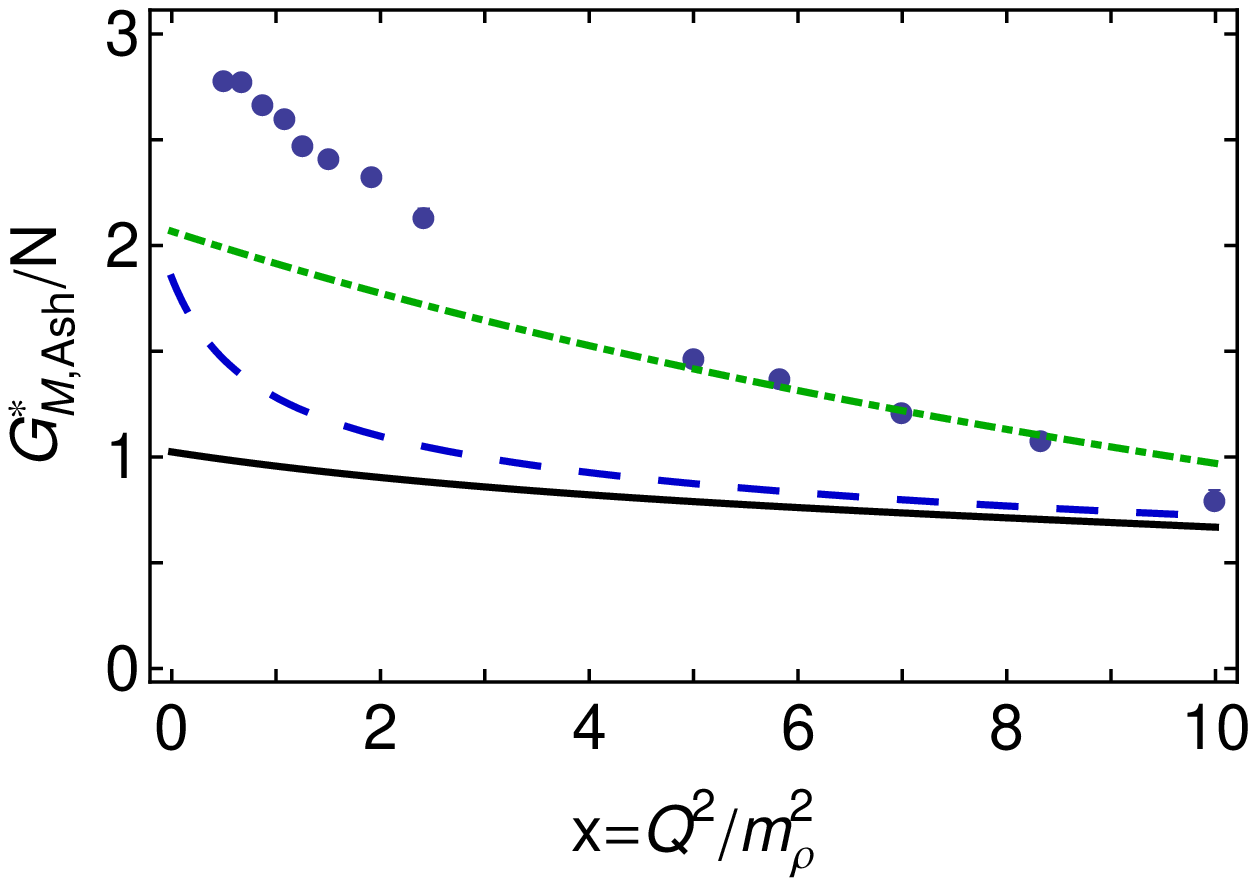}
\end{centering}
\caption{\label{figGMast}
\emph{Upper panel}.  $G_{M}^{\ast}(Q^2)$: contact-interaction result (solid curve); ameliorated result (dashed), explained around Eq.\,\protect\eqref{correction}; SL-model dressed-quark-core result \protect\cite{JuliaDiaz:2006xt} (dotted); and data from Refs.\,\protect\cite{Beringer:1900zz,Bartel:1968tw,Stein:1975yy,Sparveris:2004jn,Stave:2008aa,Aznauryan:2009mx}, whose errors are commensurate with the point size.
(N.B.\ Our formulation of the contact interaction produces Faddeev amplitudes that are independent of relative momentum, hence $G_M^\ast(Q^2)$ is hard.)
\emph{Lower panel}. $\mu_n G_{M,Ash}^{\ast}(Q^2)/N(Q^2)$: contact interaction (solid) and ameliorated result (dashed), both obtained with $N(Q^2)=G_M^n(Q^2)$.  Also, empirical results \cite{Aznauryan:2009mx} for $G_{M,Ash}^{\ast}/N_D(Q^2)$, where $1/N_D(Q^2)=[1 + Q^2/\Lambda^2]^2$, $\Lambda=0.71\,$GeV, and SL-model's dressed-quark-core result for this ratio \protect\cite{JuliaDiaz:2006xt} (dotted).}
\end{figure}

In recognition of both this defect and the general expectation that a comparison with experiment should be sensible, we subsequently provide two sets of results.  Namely, unameliorated predictions of the CI plus results obtained with two corrections: we rescale the axial($\Delta$)-scalar($N$) diagram using the factor
\begin{equation}
\label{correction}
1+ g^{as}_{aa}/[1+Q^2/m_\rho^2]\,,
\end{equation}
with $g^{as}_{aa}=4.3$, so that its contribution to $G_M^{\ast p}(0)$ matches that of the axial($\Delta$)-axial($N$) term; and incorporate a dressed-quark anomalous magnetic moment, which is a predicted consequence of DCSB in QCD \cite{Chang:2010hb,Qin:2013mta} and described in App.\,C.6 of Ref.\,\cite{Wilson:2011aa}.

In Fig.\,\ref{figGMast}, upper panel, we display the $\gamma^\ast p \to \Delta^+$ magnetic form factor.  (With $\tilde \mu_{N\Delta}^\ast:= (\sqrt{m_\Delta/m_N}) G_M^{\ast N}(0)$, we have a direct result of $\tilde \mu_{N\Delta}^\ast=1.13$ and an ameliorated value of $\tilde \mu_{N\Delta}^\ast=2.04$.)  Both curves are consistent with data for $x\gtrsim 2$ but, corrected or not, are in marked disagreement at infrared momenta.  This is explained by the similarity between our ameliorated result (dashed) and the dressed-quark-core result determined using the Sato-Lee (SL) dynamical meson-exchange model \cite{JuliaDiaz:2006xt} (dotted).  The SL result supports our view that the discrepancy results from omission of meson-cloud effects in the rainbow-ladder truncation of QCD's DSEs.

\begin{figure}[t]
\begin{centering}
\includegraphics[clip,width=0.7\linewidth]{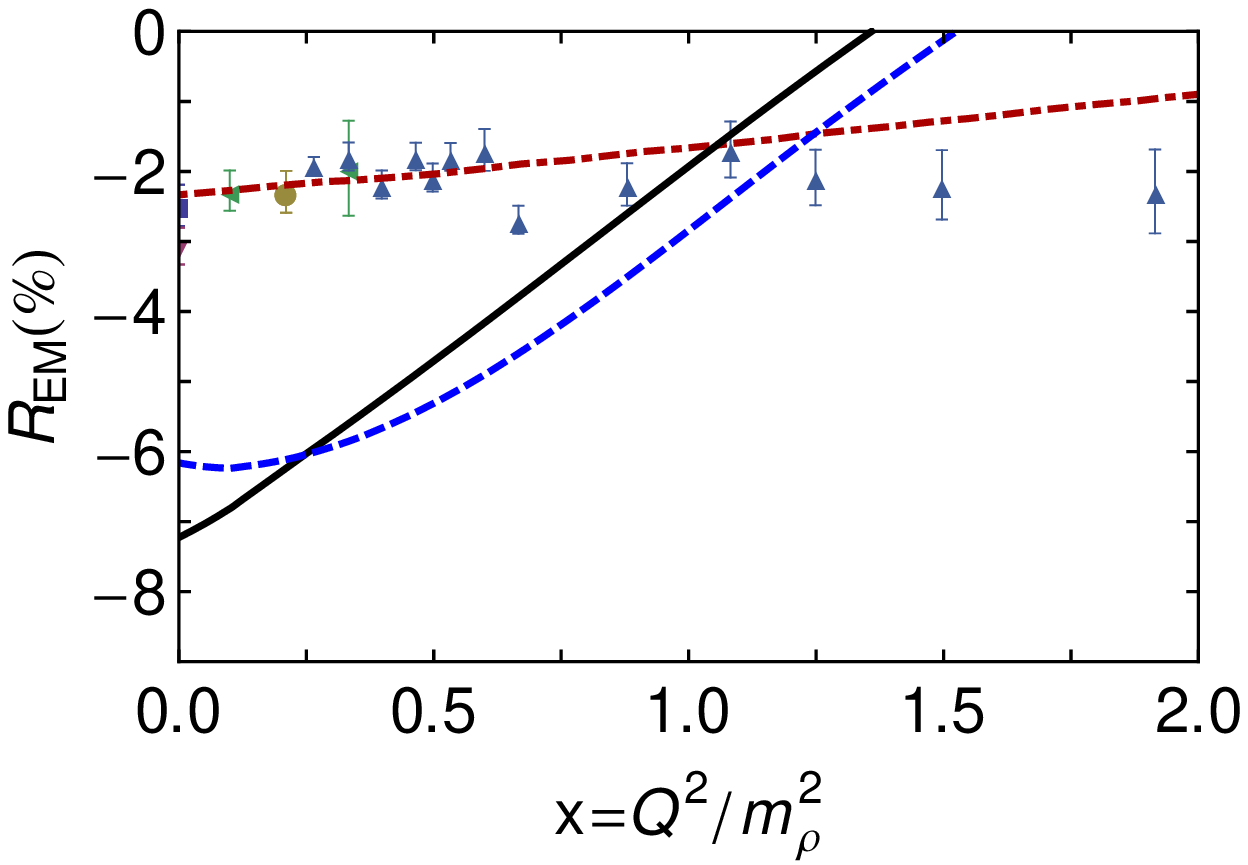}\vspace*{-1ex}

\includegraphics[clip,width=0.7\linewidth]{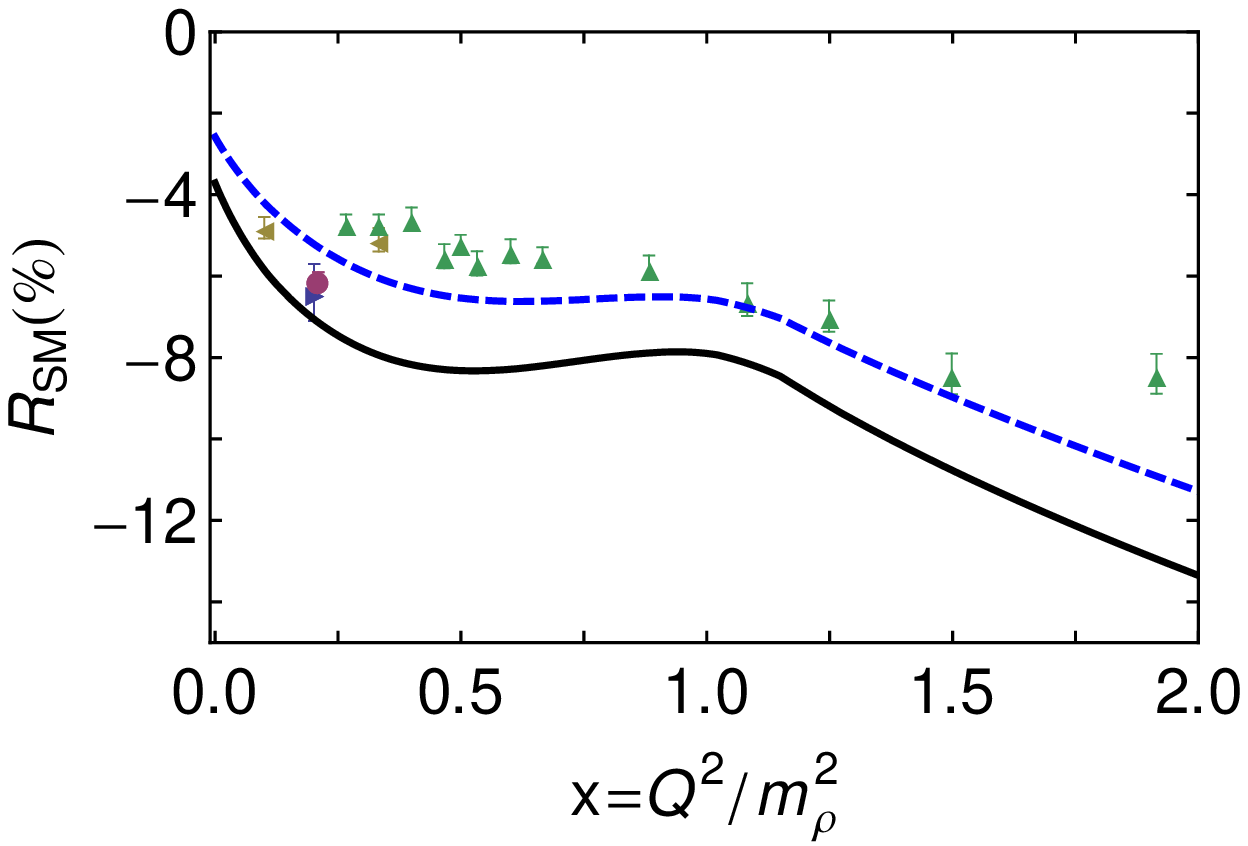}
\end{centering}
\caption{\label{figREMSM} Ratios in Eq.\,\protect\eqref{eqREMSM}.  Both panels: \emph{solid curve} -- contact-interaction result; \emph{dashed} -- ameliorated result, discussed around Eq.\,\protect\eqref{correction};
and data \protect\cite{Sparveris:2004jn,Stave:2008aa,Aznauryan:2009mx,Beck:1999ge,Pospischil:2000ad,Blanpied:2001ae}.
The \emph{dash-dot curve} in the upper panel is representative of the computation in Ref.\,\protect\cite{Eichmann:2011aa}.
(N.B.\ $G_E^\ast$, $G_C^\ast$ are small, so Ref.\,\protect\cite{JuliaDiaz:2006xt} could not reliably separate meson-cloud and dressed-quark core contributions to these ratios.)}
\end{figure}




In contrast to the upper panel of Fig.\,\ref{figGMast}, depictions of experimental data typically use the Ash form factor \cite{Ash1967165}
\begin{equation}
\label{DefineAsh}
G_{M,Ash}^{\ast}(Q^2)= G_M^{\ast}(Q^2)/[1+Q^2/t_+ ]^{1/2}.
\end{equation}
This comparison is depicted in Fig.\,\ref{figGMast}, lower panel.  (Our dressed-quark core result is quantitatively similar to the same quantity in Fig.\,3 of Ref.\,\cite{Aznauryan:2012ec}.)  Plainly, $G_{M,Ash}^{\ast}(Q^2)$ falls faster than a dipole.  Historically, many have viewed this as a conundrum.  However, as observed previously \cite{Carlson:1985mm} and elucidated herein, there is no sound reason to expect $G_{M,Ash}^{\ast}(Q^2)/G_M^n(Q^2) \approx\,$constant.  Instead, the Jones-Scadron form factor should exhibit $G_{M}^{\ast}(Q^2)/G_M^n(Q^2) \approx\,$constant.  The empirical Ash form factor falls rapidly for two reasons.  First: meson-cloud effects provide more than 30\% of the form factor for $Q^2\lesssim 2m_\rho^2$; these contributions are very soft; and hence they disappear rapidly.  Second: the additional kinematic factor $\sim 1/\sqrt{Q^2}$ in Eq.\,\eqref{DefineAsh} provides material damping for $Q^2\gtrsim 4m_\rho^2$.

In Fig.\,\ref{figREMSM} we depict the ratios
\begin{equation}
\label{eqREMSM}
\rule{-0.7em}{0ex} R_{\rm EM} = -G_E^{\ast}/G_M^{\ast}\,,\;
R_{\rm SM} = - (|\vec{Q}|/2 m_\Delta) (G_C^{\ast}/G_M^{\ast})\,,
\end{equation}
which are commonly read as measures of deformation in one or both of the hadrons involved because they are zero in $SU(6)$-symmetric constituent-quark models.  However, the ratios also measure the way in which such deformation influences the structure of the transition current.

Our results show that even a CI produces correlations between dressed-quarks within Faddeev wave-functions and related features in the current that are comparable in size with those observed empirically.  They are actually too large if axial($\Delta$)-axial($p$) contributions to the transition significantly outweigh those from axial($\Delta$)-scalar($p$) processes.  This is highlighted by the dash-dot curve in the upper panel.  That result, obtained in the same DSE truncation but with a QCD-motivated momentum-dependent interaction \cite{Eichmann:2011aa}, produces Faddeev amplitudes with a richer quark orbital angular momentum structure.  The upper panel emphasises, therefore, that $R_{\rm EM}$ is a particularly sensitive measure of orbital angular momentum correlations, both within the hadrons involved and in the excitation current.

\begin{figure}[t]
\centerline{
\includegraphics[clip,width=0.7\linewidth]{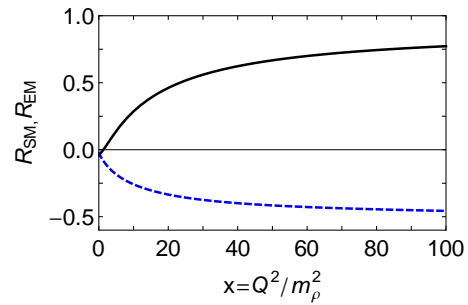}}
\caption{\label{UVREM}
$R_{\rm EM}$ (solid curve) and $R_{\rm SM}$ (dashed curve) in Eq.\,\protect\eqref{eqREMSM}, computed using the ameliorated contact interaction, discussed in connection with Eq.\,\protect\eqref{correction}.}
\end{figure}

Even though the asymptotic power-law dependence of our form factors is harder than that in QCD, one may show that helicity conservation arguments \cite{Carlson:1985mm} should apply equally to an internally-consistent symmetry-preserving treatment of a CI.  Consequently, we have
\begin{equation}
\label{eqUVREMSM}
R_{EM} \stackrel{Q^2\to\infty}{=} 1 \,,\;
R_{SM} \stackrel{Q^2\to\infty}{=} \,\mbox{\rm constant}\,.
\end{equation}
The validity of Eqs.\,\eqref{eqUVREMSM} may be read from Fig.\,\ref{UVREM}.  On one hand, it is plain that truly asymptotic $Q^2$ is required before the predictions are realised.  On the other hand, they \emph{are} apparent.  Importantly, $G_E^\ast(Q^2)$ does possess a zero (at an empirically accessible momentum) and thereafter $R_{\rm EM}\to 1$.  Moreover, $R_{\rm SM}\to\,$constant.
(N.B.\ The curve we display contains the $\ln^2 Q^2$-growth expected in QCD \cite{Idilbi:2003wj} but it is not a prominent feature.)
Since it is relative damping associated with helicity flips that yields Eqs.\,\eqref{eqUVREMSM}, with the $Q^2$-dependence of the leading amplitude being less important, it is plausible that the pattern evident herein is also that to be anticipated in QCD.

\smallskip

\hspace*{-\parindent}\textbf{IV.~Epilogue}.
We explained and illustrated that the Ash form factor connected with the $\gamma^\ast N \to \Delta$ transition should fall faster than the neutron's magnetic form factor, which is a dipole in QCD.  
In addition, we showed that the quadrupole ratios associated with this transition are a sensitive measure of quark orbital angular momentum within the nucleon and $\Delta$.  In Faddeev equation studies of baryons, this is commonly associated with the presence of strong diquark correlations.
Finally, direct calculation revealed that predictions for the asymptotic behaviour of these quadrupole ratios, which follow from considerations associated with helicity conservation, are valid, although only at truly large momentum transfers.

\smallskip

\hspace*{-\parindent}\textbf{Appendix}.
The diagrams in Fig.\,\ref{fig:Transitioncurrent} can all be expressed in the form of Eq.\,\eqref{GammaTransition}. Assuming isospin symmetry, Diagram~1 is ${\cal J}^{1N}_{\mu\alpha} = {\cal J}^{1N_1}_{\mu\alpha} + {\cal J}^{1N_0}_{\mu\alpha}$ with
\begin{subequations}
\begin{eqnarray}
%
{\cal J}^{1N_1}_{\mu\alpha} & = & (\sqrt{2}/3) \, e_{N_1} \, d^0 \,
[a_1^{0}\, I_{\mu\alpha,1}^{1\{qq\}} + a_2^{0}\, I_{\mu\alpha,2}^{1\{qq\}}]\,,
\\
{\cal J}^{1N_0}_{\mu\alpha} & = & -(\sqrt{2}/3) e_{N_0} \, d^0 \,
[a_1^{0}\, I_{\mu\alpha,1}^{1\{qq\}} + a_2^{0}\, I_{\mu\alpha,2}^{1\{qq\}} ]\,,
\end{eqnarray}
\end{subequations}
where:
$e_{p_1}=e_d=e_{n_0}$, $e_{n_1}=e_u=e_{p_0}$; and $d^0$ and $\{a^0_k,k=1,2\}$ are, respectively, canonically normalised Faddeev amplitudes for the $\Delta$ and nucleon, computed in Ref.\,\cite{Wilson:2011aa}.  In these expressions,
\begin{equation}
I_{\mu\alpha,k}^{1\{qq\}} =
\int_\ell S(\ell_f) \, i\gamma_\mu^T P_T(Q^2)\, S(\ell^+_i)\, M_{k\beta}\,\Delta^{1^+}_{\alpha\beta}(-\ell)\,,
\end{equation}
where $\int_\ell = \int\frac{d^4\ell}{(2\pi)^4}$; $\ell^\pm_{(i,f)} = \ell \pm P_{i,f}$; and $M_{1\beta}=\gamma_5\gamma_\beta$, $M_{2\beta}=\gamma_5 (\hat P_i)_\beta$.  The dressed-quark propagator, $S(p)$, and the axial-vector diquark propagator, $\Delta^{1+}_{\alpha\beta}$, are described in Ref.\,\cite{Wilson:2011aa}; whereas the dressing factor for the quark-photon vertex, $P_T(Q^2)$, is explained in Ref.\,\cite{Roberts:2011wy}.

Diagram~2 may be expressed: ${\cal J}^{2N}_{\mu\alpha} = {\cal J}^{2N_1}_{\mu\alpha} + {\cal J}^{2N_0}_{\mu\alpha}$,
\begin{subequations}
\begin{eqnarray}
%
{\cal J}^{2N_1}_{\mu\alpha} & = & (\sqrt{2}/3) \, e^{\{qq\}}_{N_1} \, d^0 \,
[a_1^{0}\, I_{\mu\alpha,1}^{2\{qq\}} + a_2^{0}\, I_{\mu\alpha,2}^{2\{qq\}}]\,,\;\;
\\
{\cal J}^{2N_0}_{\mu\alpha} & = & -(\sqrt{2}/3) e^{\{qq\}}_{N_0} \, d^0 \,
[a_1^{0}\, I_{\mu\alpha,1}^{2\{qq\}} + a_2^{0}\, I_{\mu\alpha,2}^{2\{qq\}}]\,,\;\;
\end{eqnarray}
\end{subequations}
$e^{\{qq\}}_{p_1} = 2 e_u$, $e^{\{qq\}}_{n_1} = 2 e_d$, $e^{\{qq\}}_{p_0}=e_u+e_d=e^{\{qq\}}_{n_0}$.  Here,
\begin{equation}
 I_{\mu\alpha,k}^{2\{qq\}} =
 \int_\ell \, S(\ell)\, \Delta_{\alpha\rho}^{1^{+}}(-l^-_{f})
 \Gamma_{\mu,\rho\sigma}^{1^{+}} \Delta_{\sigma\beta}^{1^{+}}(-l^-_{i}) M_{k\beta},
\end{equation}
wherein $\Gamma_{\mu,\rho\sigma}^{1^{+}}(-l^-_{f},-l^-_{i}) $ is the dressed--photon--pseudovector-diquark vertex (see Sec.\,IV.B, Ref.\,\cite{Roberts:2011wy}).

Diagram~3, the bottom image in Fig.\,\ref{fig:Transitioncurrent}, describes an electromagnetically-induced $0^+-1^+$ diquark transition:
${\cal J}^{3p}_{\mu\alpha} = -{\cal J}^{3n}_{\mu\alpha}
= d^0 s_N \, (e_u+e_d)\,I_{\mu\alpha}^{3}$,
where
\begin{equation}
I_{\mu\alpha}^{3} = \int_\ell S(l) \, \Delta_{\alpha\rho}^{1^{+}}(-l^-_{f})\,
i\Gamma^{10}_{\rho\mu}(-l^-_{f},-l^-_{i}) \Delta^{0^{+}}(-l^-_{i}).
\end{equation}
The transition form factor $\Gamma^{10}_{\rho\mu}(\ell_2,\ell_1) = \Gamma^{01}_{\mu\rho}(\ell_1,\ell_2)$ is detailed in Sec.\,IV.C of Ref.\,\cite{Roberts:2011wy}.


At this point, we follow the steps explained in Apps.\,C.5 and D of Ref.\,\cite{Wilson:2011aa} in order to arrive at a concrete expression for $\Gamma_{\alpha\mu}(K,Q)$ in Eq.\,\eqref{eq:Gamma2Transition} and subsequently, via Eqs.\,\eqref{contractions}, numerical results for the three Poincar\'e-invariant transition form factors in Eq.\,\eqref{eq:Gamma2Transition}.

Before producing final results for the transition form factors, one must also calculate nucleon and $\Delta$ elastic form factors within the same framework in order, at least, to compute the canonical normalisation constants for the nucleon and $\Delta$ Faddeev amplitudes.  (These constants ensure unit electric charge for the proton and $\Delta^+$ \cite{Oettel:1999gc}.)  This has already been done for the nucleon \cite{Wilson:2011aa}; and for the $\Delta$, we trace the pattern in Ref.\,\cite{Nicmorus:2010sd}, adapted as necessary, following Ref.\,\cite{Wilson:2011aa}, to the CI.

One must proceed carefully with that calculation, however.  Using the Ward-Green-Takahashi identities for the quark-photon and quark-diquark vertices, one can show that, at $Q=0$, Diagrams~1 and 2 must be equal.  Computationally, this is ensured by any $O(4)$- and translationally-invariant regularisation scheme.  Whilst both are formally a property of our treatment of the CI, the latter is practically broken by the final step of introducing infrared and ultraviolet mass-scales in the proper-time regularisation of integrals.  The effect is to produce a small mismatch between these diagrams at $Q=0$.
The weakness can be traced to quadratic divergences that arise through integrals such as
\begin{eqnarray}
\nonumber
&&
\int_\ell
\frac{1}{[\ell^2+\omega]^2} \{(K\cdot \ell)^2,(Q\cdot \ell)^2, (K\cdot \ell) (Q\cdot \ell)\}\\
&=& 
\int_\ell \frac{\ell^2}{[\ell^2+\omega]^2}\frac{1}{4}
\{K^2,Q^2 , K\cdot Q \}\,, \label{eq:1on4}
\end{eqnarray}
and analogous integrals with a quartic divergence.  As explained elsewhere \cite{Chen:2012txa}, the weakness can be ameliorated via a simple expedient: in Eq.\,\eqref{eq:1on4}, replace $1/4 \to \theta = 1.874 (1/4)$; and in results for those integrals with a quartic divergence, replace the usual factor of $1/24$ by $\theta^2/24 $.

We arrive in this way at, \emph{inter alia} \cite{Segovia:2013uga}, magnetic moments for the $\Delta$-resonances, measured in nuclear magnetons:
$\mu_{\Delta^q} = 3.1q$, $q=+2,+1,0,-1$.  These values may be compared with lattice-QCD results \cite{Cloet:2003jm}: $\mu_{\Delta^+}=2.5$; and DSE results obtained using a QCD-motivated momentum-dependent interaction: $\mu_{\Delta^+} = 2.7 q$.
%

\smallskip

\noindent\textbf{Acknowledgments}.
We are grateful for valuable input from R.~Gothe and V.~Mokeev.
C.~Chen acknowledges the support of the China Scholarship Council (file no.\,2010634019).  This work was otherwise supported by
U.\,S.\ Department of Energy, Office of Nuclear Physics, contract no.~DE-AC02-06CH11357.
%


\end{document}